# Gaussian Process Latent Class Choice Models

## A Preprint

**Georges Sfeir**

Ph.D. Candidate, Civil and Environmental Engineering
American University of Beirut, Beirut, Lebanon, Riad el Solh 1107 2020
E-mail: gms12@mail.aub.edu

**Filipe Rodrigues**

Associate Professor, DTU Management Engineering, Transport DTU
Technical University of Denmark, Kgs. Lyngby, Denmark, 2800
E-mail: rodr@dtu.dk

**Maya Abou-Zeid**

Associate Professor of Civil and Environmental Engineering
American University of Beirut, Beirut, Lebanon, Riad el Solh 1107 2020
E-mail: ma202@aub.edu.lb

## ABSTRACT

We present a Gaussian Process – Latent Class Choice Model (GP-LCCM) to integrate a non-parametric class of probabilistic machine learning within discrete choice models (DCMs). Gaussian Processes (GPs) are kernel-based algorithms that incorporate expert knowledge by assuming priors over latent functions rather than priors over parameters, which makes them more flexible in addressing nonlinear problems. By integrating a Gaussian Process within a LCCM structure, we aim at improving discrete representations of unobserved heterogeneity. The proposed model would assign individuals probabilistically to behaviorally homogeneous clusters (latent classes) using GPs and simultaneously estimate class-specific choice models by relying on random utility models. Furthermore, we derive and implement an Expectation-Maximization (EM) algorithm to jointly estimate/infer the hyperparameters of the GP kernel function and the class-specific choice parameters by relying on a Laplace approximation and gradient-based numerical optimization methods, respectively. The model is tested on two different mode choice applications and compared against different LCCM benchmarks. Results show that GP-LCCM allows for a more complex and flexible representation of heterogeneity and improves both in-sample fit and out-of-sample predictive power. Moreover, behavioral and economic interpretability is maintained at the class-specific choice model level while local interpretation of the latent classes can still be achieved, although the non-parametric characteristic of GPs lessens the transparency of the model.

# 1 INTRODUCTION

For decades, Discrete Choice Models (DCMs) have been used to model and forecast human decision making in different fields such as marketing, economics, psychology and especially transportation. Recently, due to the availability of advanced computer hardware and big data from mobile phones, social networks, and Internet-of-things, several studies have tried to apply machine learning (ML) algorithms to different transportation research areas (e.g. traffic control, incident detection, traffic forecasting, prediction of transportation modes from raw GPS data, demand modelling). Machine learning algorithms are non/parametric approaches that learn from the data without imposing strict statistical assumptions and can be used to capture complex patterns. Such methods are able to achieve high classification and prediction accuracy (e.g., prediction of transport modes used), but unlike discrete choice models, cannot be used to directly infer marginal effects and economic indicators such as elasticities, willingness to pay, and consumer welfare measures, which are important measures used in transportation policy and project evaluation (Hillel et al., 2020; Paredes et al., 2017). On the other hand, DCMs might not guarantee high prediction accuracy (Sifringer et al., 2020).

Lately, Gaussian Processes (GPs), a popular non-parametric class of probabilistic machine learning (Rasmussen and Williams, 2006), have been receiving growing attention and are being applied to many regression and classification applications within transportation such as travel time prediction (Idé and Kato, 2009; Rodrigues et al., 2016), crowdsourced traffic data (Rodrigues et al., 2019; Rodrigues and Pereira, 2018), congestion and routing models (Liu et al., 2013), traffic volume forecasting (Xie et al., 2010), censored demand modeling (Gammelli et al., 2020), etc. Compared to other machine learning methods, Gaussian Processes are considered more attractive due to their flexible non-parametric nature and their formulation in a full Bayesian framework, something that guarantees probabilistic interpretation of the model outputs (Mackay, 1997). Moreover, GPs are also kernel-based algorithms that assume priors over latent functions rather than priors directly over parameters, which makes GPs very powerful in addressing difficult nonlinear regression and classifications problems (Rasmussen and Williams, 2006; Seeger, 2004).

Returning to discrete choice models, several advanced models have been developed to model complex choice situations and to tackle different problems such as heterogeneity in behavior and endogeneity. However, the question of how best to model unobserved heterogeneity remains one of the most active research areas within demand modeling (Vij and Krueger, 2017). The mixed logit family, where choice probabilities are a weighted average of standard logit probabilities over some mixing distributions (Train, 2009), is by far the most popular approach for capturing heterogeneity. The literature is rich with studies and information on different types of mixing distributions such as continuous, discrete, parametric and nonparametric. For a recent review that comprehensively covers the different types of mixing distributions in addition to their advantages and disadvantages, readers may refer to Yuan et al. (2015). While the ongoing debate between continuous and discrete representations of heterogeneity is beyond the scope of this paper, this study focuses on the discrete nonparametric category of mixing distributions. The Latent Class Choice Model (LCCM) remains the most popular and well-established example of discrete nonparametric mixing distributions and can be described as a mixed logit model with a finite mixing distribution (Train, 2008; Yuan et al., 2015). LCCM is a random-utility model that is used whenever the modeler hypothesizes that the unobserved heterogeneity can be represented through discrete segments (or latent classes) of people that differ behaviorally from each other due to varying tastes, different decision protocols adopted by individuals and/or different choice sets considered by each individual (Gopinath, 1995). LCCM consists of two sub-models, a class membership model that formulates the probability of an individual belonging to a specific segment/class and a class-specific choice model that estimates the choice probabilities. However, the linear-in-parameters utility specification of the latent classes may oversimplify and underestimate the extent of behavioral heterogeneity within a population.

The objective of this study is to integrate Gaussian Processes into the LCCM structure to allow for more complex and flexible discrete representation of heterogeneity and as a result to improve the overall model fit and prediction accuracy. We propose a GP-LCCM framework that makes use of Gaussian Processes to replace the class membership component of the traditional LCCM. The proposed model would rely on GPs as a non-parametric component to probabilistically divide the population into behaviorally homogenous classes while simultaneously relying on random utility models to develop class-specific choice models. We derive and implement an Expectation-Maximization (EM) algorithm for training a Gaussian Process classification approach as a clustering tool while concurrently learning the parameter estimates of the class-specific choice models. By doing so, we contribute to the discrete choice modeling literature by formulating, to the authors' knowledge, the first Gaussian Process choice model within an LCMM framework, thereby allowing for more modeling flexibility and higher prediction accuracy. We also develop a Gaussian Process model for clustering by incorporating the Laplace approximation approach (Williams and Barber, 1998), which is used for Gaussian process classification problems, in an iterative EM algorithm. To illustrate the proposed model, we apply it to two different mode choice applications and compare the results to different discrete choice models. As empirical evidence later shows, the proposed model improves model goodness-of-fit, generalization performance and heterogeneity representation, all while ensuring that economic indicators can be easily extracted.

The remainder of this paper is organized as follows. Section 2 reviews the literature on discrete choice models, machine learning and Gaussian Processes. Section 3 presents the mathematical formulation of the proposed GP-LCCM. Section 4 presents and compares the estimation results of two different case studies. Section 5 summarizes the findings and discusses future extensions of this work.

## 2 LITERATURE

Section 2.1 reviews the problem of representing heterogeneity in discrete choice models while Section 2.2 reviews studies that have used machine learning in mode choice modeling and discusses the use of Gaussian Processes. Throughout, we discuss the advantages and disadvantages of each field in order to motivate the need for the model proposed by this study (Section 2.3).

### 2.1 DCM and Heterogeneity

Representing unobserved heterogeneity in consumers' behavior remains a popular topic in the discrete choice modeling literature. Several studies have found that misrepresenting heterogeneity in the choice process can result in unreliable parameter estimates, false conclusions, and misleading forecasts (Gopinath, 1995; Vij et al., 2013). Mixed logit remains, by far, the most popular framework for representing random heterogeneity. Such models can mimic any random utility model to high levels of accuracy by assuming parametric distribution(s) with predefined forms (e.g. normal or lognormal) for the random parameters and/or random components (McFadden and Train, 2000). However, such models are constrained by the predefined forms of the parametric mixing distribution(s) and therefore they may be subject to misspecification (Vij and Krueger, 2017; Yuan et al., 2015). In addition, choosing the true parametric shape is usually a complicated and computationally expensive task that requires the estimation of several models with different parametric distributions and an exhaustive comparison based on statistical goodness-of-fit measures and model interpretability (Train, 2008; Vij and Krueger, 2017). In contrast, nonparametric distributions overcome the aforementioned constraints since they do not have predefined forms, meaning more flexibility can be guaranteed (Yuan et al., 2015). Latent Class Choice Model (LCCM) is the most popular nonparametric distribution model and is usually adopted when the analyst hypothesizes that the unobserved heterogeneity can be represented through discrete constructs such as different decision protocols used by individuals, segments of the population with varying tastes, and choice sets considered which may vary from one individual to another (Gopinath, 1995). In general, the flexibility of LCCM increases with the number of classes. However, the computational complexity grows rapidly since the number of parameters increases with the number of classes as well (Yuan et al., 2015). Consequently, the computational burden that precludes the estimation of LCCMs with a high number of classes in addition to the linear-in-parameters specification of the latent classes may generate in practice simpler (less flexible) models than the LCCM framework can offer (Vij and Krueger, 2017).

Some studies have implemented mixture of distributions approaches in an attempt to overcome the limitations of both Mixed logit and LCCM. For instance, Bujosa et al. (2010) and Green and Hensher (2013) integrated random taste coefficients in the class-specific utilities of LCCM to account for within-class heterogeneity. The two proposed models demonstrated better goodness-of-fit measures than both LCCM and Mixed logit. However, the applications were limited to the introduction of one random taste coefficient to an LCCM with two classes. Krueger et al. (2018) developed a Dirichlet Process mixture model that does not require to determine the number of mixture components in advance. However, the use of Dirichlet Process affects the overall behavioral interpretability of the proposed model since it leads to unstructured representations of heterogeneity. Train (2016) used logit specifications to model the mixing distribution of random parameters. The proposed model has the capability to mimic any mixing distribution with greater flexibility. However, it requires a manual search for the optimal utility specification of the random parameters.

## 2.2 Machine Learning in Choice Modeling

In recent years, the use of machine learning techniques has increased exponentially. Such methods are being applied to problems from different fields (speech processing, computational biology, finance, robotics, computer vision, natural language processing, etc.). As for transportation, researchers have been exploring the feasibility of applying machine learning techniques to different transportation research areas such as traffic control (Abdulhai et al., 2003; Bingham, 2001; Srinivasan et al., 2006), incident detection (Jin et al., 2002; Srinivasan et al., 2004; Wang et al., 2008), traffic forecasting (Deshpande and Bajaj, 2017; Hong et al., 2011; Ma et al., 2018; Stathopoulos et al., 2008; Vlahogianni et al., 2008) and prediction of transportation modes from raw GPS data and/or mobile phone sensors such as accelerometers and gyroscopes (Dabiri and Heaslip, 2018; Gonzalez et al., 2010; Jahangiri and Rakha, 2015, 2014; Zhu et al., 2017).

Moreover, supervised machine learning techniques are increasingly being used in mode choice modeling as alternative methods to traditional econometric models. While most of the contrast studies have shown that ML outperforms DCM in terms of prediction accuracy (Andrade et al., 2006; Cantarella and de Luca, 2005; Lee et al., 2018; Nijkamp et al., 1996; Sekhar et al., 2016;

Tang et al., 2015; Wang et al., 2020a; Xian-Yu, 2011; Xie et al., 2003), some studies have shown no clear advantage of machine learning (Hensher and Ton, 2000; Sayed and Razavi, 2000). However, econometricians and transportation researchers are still relying on traditional econometric models instead of machine learning techniques. This may be due to the differences in the underlying philosophy and goals of the two approaches. Machine learning models can be referred to as "predictive" models that target high classification and prediction accuracy at the expense of interpretability, although some studies have shown that different techniques can be applied to extract some behavioral indicators (e.g. Andrade et al., 2006; Ding et al., 2018; Wang et al., 2020b). On the other hand, traditional discrete choice models are known as "explanatory" models that assume parametric relationships between the utility of each alternative and its potential attributes. They can be used to directly infer marginal effects and economic indicators such as elasticities, willingness to pay, and consumer welfare measures. However, these explanatory models might not guarantee high prediction accuracy. Moreover, traditional DCMs are rooted in microeconomic theories of human decision-making behavior (Bierlaire and Lurkin, 2017). It is believed that this connection is the main reason econometricians have heavily relied on discrete choice models and specifically the Multinomial Logit (MNL) formulation of McFadden (Brathwaite et al., 2017; McFadden, 2001). It is also believed that the main reason that kept econometricians from trusting machine learning is the missing link with economic theories (Brathwaite et al., 2017). Nonetheless, some studies have tried to connect the two fields through hybrid frameworks. For instance, a two-stage sequential logit-Artificial Neural Networks (ANN) framework for choice modeling has been proposed by Gazder and Ratrout (2016). Results showed that the proposed model improves the generalization performance (prediction accuracy) of logit models in multinomial choice situations while logit models have marginally higher prediction accuracies in binary choice situations. Sifringer et al. (2018) also developed a two-stage sequential model by adding an extra term, estimated through an ANN, to the utilities of the logit model. This approach improved the log-likelihood of the simple logit model by around 15% without weakening the statistical significance of the logit parameters. Another choice model that combines neural networks and random utility models (Wong and Farooq, 2019) has been developed by using the concept of residual learning within a neural network architecture to allow for training of deep neural networks and as such identifying complex sources of unobserved heterogeneity. Moreover, Wong et al. (2018) estimated latent variables without attitudinal indicators, as in Integrated Choice and Latent Variable models (ICLV), by constructing a restricted Boltzmann machine - a generative non-parametric ML approach.

## 2.3 This Study in Context

Most of the previous studies have focused on applying supervised machine learning to classifications tasks such as mode choice modeling or on combining machine learning techniques (mostly Neural Networks) and discrete choice models in sequential or simultaneous framework. Furthermore, the majority of studies that have used machine learning methods as alternatives to discrete choice models for mode choice modeling have mostly focused on supervised applications and prediction accuracy (Karlaftis and Vlahogianni, 2011; Wong and Farooq, 2019) often at the expense of economic interpretability due to the disconnection of such techniques with economic principles and theories (Brathwaite et al., 2017). Some recent studies have shown though that machine learning techniques can provide practical economic information (Brathwaite et al., 2017; Wang et al., 2020b). In this study, we aim at using Gaussian Processes for clustering tasks within the famous LCCM econometric framework. As previously mentioned, most of the studies on machine learning for mode choice modeling have been related to classification[1] tasks while clustering[2] techniques have not yet been rigorously addressed in the literature, although some studies have addressed this clustering aspect of machine learning. For instance, Han (2019) developed a nonlinear-LCCM by using neural networks to specify the class membership model. The proposed model with 8 latent classes outperformed the best LCCM with 6 latent classes in terms of prediction accuracy. However, the nonlinear-LCCM is less transparent and loses interpretability at the latent class level due to the "black-box" nature of neural networks. Moreover, Sfeir et al. (2020b) developed a flexible LCCM (GBM-LCCM) by using Gaussian-Bernoulli Mixture Models to probabilistically cluster individuals to different homogenous classes within a LCCM framework. The results showed that the model is capable of improving the out-of-sample prediction accuracy in addition to capturing more complex taste heterogeneity without weakening the behavioral and economic interpretability of the choice models. However, the Gaussian-Bernoulli Mixture Model assumes that continuous and categorical variables are uncorrelated and as such the relations between the different variables used for clustering cannot be well determined. Moreover, Gaussian-Bernoulli Mixture Models as well as simple neural networks are parametric models. Such models assume specific functional form for the distribution (or mapping function) and as such have predefined numbers of parameters that once learned (estimated) would be used for predictions. On one hand, parametric assumptions make the learning process (estimation) easier, faster and in less need of training data to learn the parameters. On the other hand, parametric models are constrained to the functional forms they assume, which restricts the flexibility of the learning process and might lead to poor generalization or prediction accuracies (Ghahramani, 2015). An alternative approach to improve the flexibility and generalization performance is to rely on non-parametric machine learning algorithms. Such methods are data-driven, do not assume predefined functional forms and consequently are free to learn any functional form from the training data. These methods still contain parameters to control the complexity of the model rather than the functional form of the distribution (Bishop, 2006). Gaussian Process is one such method that avoids simple parametric assumptions and provides a fully Bayesian framework for modeling (Rasmussen and Williams, 2006). These characteristics make GPs very attractive for modeling uncertainties and complex nonlinear problems. Moreover, Gaussian Process can be mathematically equivalent to neural networks with very large number of hidden units (Neal, 1996). However, Gaussian Processes are generally easier to handle since the estimation of a neural network is usually complicated by the fact that the optimization problem might have several local optima while the posterior of the Gaussian Process for regression and classification is convex (Rasmussen and Williams, 2006).

## 3 MODEL FRAMEWORK AND FORMULATION

This study develops a Gaussian Process – Latent Class Choice Model (GP-LCCM) by incorporating a Gaussian Process within a discrete choice model framework to improve the representation of heterogeneity while maintaining the econometric interpretability of random utility models. We start by presenting the LCCM formulation (Section 3.1). Next, we describe the Gaussian Process formulation for binary classification (Section 3.2) and present different kernel functions (Section 3.3). Finally, we combine the three concepts (LCCM, GP and kernels) to define the Gaussian Process – Latent Class Choice Model (GP-LCCM) and derive an Expectation-Maximization (EM) algorithm for estimation (Section 3.4).

---

[1] Classification algorithms are a sub-category of supervised ML which makes use of labeled data (data that includes both independent variables and dependent variables or labels) to build a classifier (learn a function) that would predict discrete labels.

[2] Clustering methods are a sub-category of unsupervised ML which are used to identify underlying patterns or discover clusters (groups or classes) of similar characteristics within an unlabeled data (data that only has independent variables).

## 3.1 Latent Class Choice Model

LCCM consists of two components, a class membership model and a class-specific choice model. The class membership model estimates the probability that a decision-maker belongs to a specific class, typically as a function of his/her characteristics. The utility of belonging to latent class $k$ for decision-maker $n$ is defined as follows:

$$U_{nk} = S'_n \gamma_k + \nu_{nk}, \tag{1}$$

with $S_n$ a vector of socio-economic/demographic variables of decision-maker $n$ including a constant, $\gamma_k$ the corresponding vector of unknown parameters, and $\nu_{nk}$ a random disturbance term that is assumed to follow an independently and identically distributed ($iid$) Extreme Value Type I distribution over decision-makers and classes.

The class membership probability for decision-maker $n$ and latent class $k$ is then defined as follows:

$$P(q_{nk} = 1|S_n, \gamma_k) = \frac{e^{S'_n \gamma_k}}{\sum_{k'=1}^{K} e^{S'_n \gamma_{k'}}}, \tag{2}$$

with $q_{nk}$ equal to 1 if decision-maker $n$ belongs to latent class $k$ and 0 otherwise.

The class-specific choice model formulates the probability of selecting a specific alternative, conditional on the class membership assignments, as a function of the exogenous attributes of the alternatives. As such, the utility of decision-maker $n$ selecting alternative $j$ during time period / choice occasion $t$, conditional on him/her belonging to class $k$, is specified as:

$$U_{njt|k} = X'_{njt} \beta_k + \varepsilon_{njt|k}, \tag{3}$$

where $X_{njt}$ is a vector of exogenous attributes related to alternative $j$ during time period $t$ and including a constant, $\beta_k$ is the corresponding vector of unknown parameters, and $\varepsilon_{njt|k}$ is a random disturbance term that is assumed to follow an $iid$ Extreme Value Type I distribution over alternatives, decision-makers and classes.

Conditioned on class $k$, the probability of decision-maker $n$ selecting an alternative $j$ in time period $t$ can then be written as follows:

$$P(y_{njt} = 1|X_{njt}, q_{nk} = 1, \beta_k) = \frac{e^{V_{njt|k}}}{\sum_{j'=1}^{J} e^{V_{nj't|k}}}, \tag{4}$$

with $J$ being the total number of alternatives.

Assuming that the conditional choice probabilities (Equation 4) for decision-maker $n$ over all time periods $T_n$ are conditionally independent, the conditional probability of observing a $(J \times T_n)$ matrix of choices $y_n$ can be expressed as follows:

$$P(y_n|X_n, q_{nk} = 1, \beta_k) = \prod_{t=1}^{T_n} \prod_{j=1}^{J} \left(P(y_{njt} = 1|X_{njt}, q_{nk} = 1, \beta_k)\right)^{y_{njt}}, \tag{5}$$

with $X_n$ being a matrix consisting of $J \times T_n$ vectors of $X_{njt}$, $y_n$ a $(J \times T_n)$ matrix of all choices of individual $n$ during all time periods $T_n$ and consisting of choice indicators $y_{njt}$, and $y_{njt}$ a choice indicator equal to 1 if decision-maker $n$ chooses alternative $j$ during time period $t$ and 0 otherwise.

The unconditional probability of the observed choice of decision-maker $n$ is then obtained by summing the product of the class membership probability (Equation 2) by the conditional choice probability (Equation 5) over all latent classes:

$$P(y_n) = \sum_{k=1}^{K} P(q_{nk} = 1|S_n, \gamma_k) P(y_n|X_n, q_{nk} = 1, \beta_k). \tag{6}$$

Finally, the likelihood over a sample of independent decision-makers $N$ is:

$$P(y) = \prod_{n=1}^{N} \sum_{k=1}^{K} P(q_{nk} = 1|S_n, \gamma_k) P(y_n|X_n, q_{nk} = 1, \beta_k). \tag{7}$$

### 3.2 Gaussian Process

Gaussian Processes (GPs) are a powerful and flexible probabilistic machine learning technique (Rasmussen and Williams, 2006) that instead of parameterizing the target variables (outputs) or placing priors over the unknown parameters of a predefined distribution (e.g. mean and variance of a normal distribution), define priors over latent functions directly (Mackay, 2003; Rasmussen and Williams, 2006). It can be considered as a generalization of a Gaussian distribution over a finite vector space to an infinite function space (Mackay, 2003). Therefore, a GP is specified by a mean function and a covariance function usually known as kernel.

For the sake of simplicity and without loss of generality, we consider a binary case ($K = 2$) where the training data consists of $S$, a matrix of $N$ vectors $S_n$ (vector of characteristics of decision-maker $n$ with a dimension equal to D), and $q_k$ a vector of $N$ target outputs $q_{nk}$ (class label, equal to 1 or 0). The goal is to model the posterior distribution of the target outputs by defining a prior distribution over a latent function $f$ by using a multivariate Gaussian distribution with a mean function $m(S_n)$, that represents the expected value for each latent variable $f(S_n)$, and a kernel (covariance function) $C(S_n, S_m) = cov[f(S_n), f(S_m)]$, that represents the variance between every pair of variables $f(S_n)$ and $f(S_m)$. Note that it is common to specify a GP with a zero mean function without loss of generality (Rasmussen and Williams, 2006). A GP prior is therefore specified for the function values $f$, $f \sim GP(m(Sn) = 0, C(S_n, S_m))$ such that:

$$p(f|S) = \mathcal{N}(0, C), \tag{8}$$

where $f$ is a vector of $N$ latent variable values $f_n$, $S$ is a matrix of $N$ vectors of $S_n$ and $C$ is a ($N \times N$) covariance matrix defined by a covariance function (or kernel) such that $C_{n,m} = C(S_n, S_m)$. Note that $C$ could be a group of $D$ ($N \times N$) matrices in case an Automatic Relevance Determination (ARD) covariance function is used (Refer to section 3.3).

The next step is to specify an appropriate likelihood or link function for the classes to obtain a probabilistic classification since the target outputs are discrete (0 or 1). The link function could be a sigmoid function or a cumulative density function of a standard normal distribution. We make use of a sigmoid function as follows:

$$P(q_{nk}|f_n) = \frac{1}{1 + exp(-f_n)}. \tag{9}$$

Then, the posterior over $f$ can be determined using Bayes' theorem as follows:

$$P(f_n|q_{nk}, S_n) = \frac{P(q_{nk}|f_n)P(f_n|S_n)}{P(q_{nk}|S_n)}. \tag{10}$$

The combination of a Gaussian process prior with a non-Gaussian link function results in a non-Gaussian posterior that is analytically intractable. Nevertheless, the posterior can be approximated by a GP using different approximation techniques such as Markov Chain Monte Carlo (MCMC) (Neal, 1999), Variational Inference (VI) (Gibbs and Mackay, 2000) and Expectation Propagation (EP) (Minka, 2001; Opper and Winther, 2000). In this study, we make use of the Laplace approximation (Bishop, 2006; Rasmussen and Williams, 2006; Williams and Barber, 1998) which approximates the posterior with a Gaussian by taking a second-order Taylor expansion of the logarithm of the posterior around its maximum. For more details about the Laplace approximation, readers may refer to (Rasmussen and Williams, 2006, sec. 3.4) and (Bishop, 2006, sec. 6.4.6).

## 3.3 Kernels

The choice of a suitable covariance function (kernel) is a crucial step in learning a Gaussian Process that generalizes beyond the training data since the kernel can shape the distribution we wish to learn in many different ways and determine the characteristics of the fitted function (such as smoothness, periodicity, stationarity and isotropy). Different kernels or combinations (addition or multiplication) of kernels can be used to generate more complex structures and improve the GP flexibility. We present next the most common kernels and the ones that are used in our applications (Section 4).

### 3.3.1 Squared Exponential Kernel (SE) / Radial Basis Function (RBF)

The most common choice of kernel is the Squared Exponential kernel (SE), also known as Radial Basis Function (RBF), which is defined as follows:

$$k_{SE}(S_n, S_m) = \lambda^2 exp\left(\frac{r^2}{2\ell^2}\right), \tag{11}$$

where $r = |S_n - S_m|$ is the Euclidean distance between two observations $S_n$ and $S_m$ (e.g. characteristics of two individuals $n$ and $m$), $\lambda^2$ is the variance of the distance between two observations and $\ell$ is the length-scale which determines the smoothness of the kernel function and the importance of the features (independent variables). Large length-scale values mean the function values are uncorrelated and the corresponding feature(s) should be removed from the model (Rasmussen and Williams, 2006).

The SE kernel is a stationary kernel that is infinitely differentiable (mean square derivatives of all orders) and as such is very smooth. However, it is believed that such strong smoothness is unrealistic for some applications and the Matérn kernel is instead recommended (Stein, 1999).

*3.3.2 Matérn Kernel*

The Matérn kernel is a stationary kernel that can be considered as a generalization of the SE kernel. It is defined as follows:

$$k_{Matern}(S_n, S_m) = \frac{2^{1-\nu}}{\Gamma(\nu)} \left(\frac{\sqrt{2\nu}r}{\ell}\right)^\nu K_\nu\left(\frac{\sqrt{2\nu}r}{\ell}\right), \tag{12}$$

where $r = |S_n - S_m|$ is the Euclidean distance between two observations $S_n$ and $S_m$ (e.g. characteristics of two individuals $n$ and $m$), $\nu$ is a positive parameter that controls the smoothness of the function (lower values result in less smooth functions), $\ell$ is the length-scale of the kernel, $\Gamma$ is the gamma function, and $K_\nu$ is a modified Bessel function (Abramowitz and Stegun, 1965, sec. 9.6)(Mackay, 1998).

The most interesting and commonly used cases for machine learning are $\nu = 3/2$ and $\nu = 5/2$.

It is to be noted that both of the above kernels could be used with Automatic Relevance Determination (ARD) by specifying the length-scale as a vector of dimension $D$ equal to the dimension of $S_n$ (Rasmussen and Williams, 2006). Other kernel functions can be used such as periodic, exponential, radial quadratic, or piecewise polynomial, to name a few. For more details, readers may refer to (Rasmussen and Williams, 2006, sec. 4).

## 3.4 GP-LCCM

We now present the formulation of the Gaussian Process – Latent Class Choice Model (GP-LCCM). Similar to the traditional LCCM, the GP-LCCM consists of two components, a class membership model and a class-specific choice model. The former is defined as a Gaussian Process to probabilistically assign decision-makers to behaviorally homogeneous latent classes/clusters, while the latter formulates class-specific choice probabilities using typical discrete choice models (e.g. MNL). Figure 1 shows the graphical representation of the proposed model. Hatched circles represent observed variables and choices while white circles symbolize unknown/latent variables.

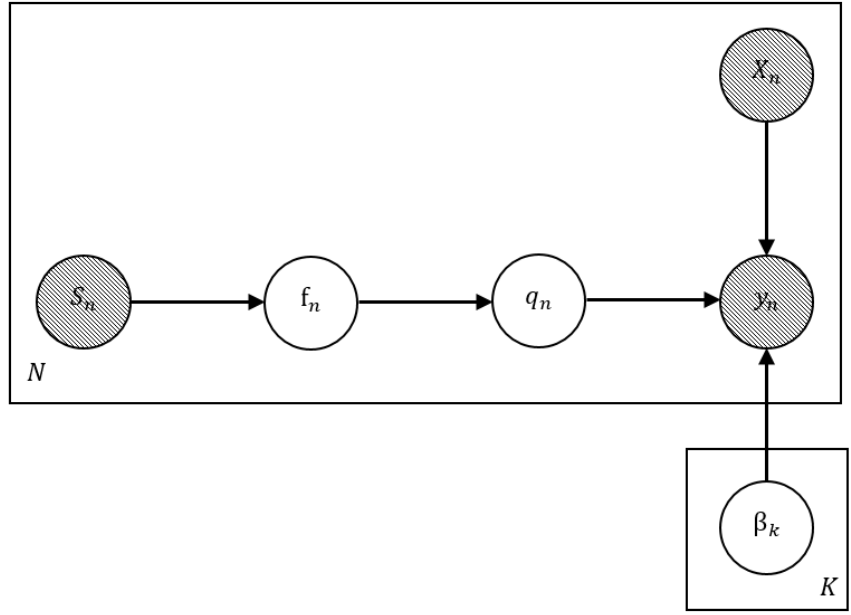

**Figure 1: Graphical representation of the proposed Gaussian Process – Latent Class Choice Model (GP-LCCM) for a set of N decision-makers and K clusters/latent classes**

*3.4.1 Proposed Model*

Given the conditional independence properties of the graphical model structure of the GP-LCCM (Figure 1), the joint probability of $f_n$, $y_n$, and $q_{nk}$ can be formulated as the product of the Gaussian prior (Equation 8 and first term on the right hand side below), the link function or the likelihood of $q_{nk}$ conditional on the latent function $f_n$ (Equation 9 and second term on the right hand side below) and the choice probability conditional on the class (Equation 5 and third term on the right hand side below), as follows:

$$P(f_n, y_n, q_{nk}|S_n, X_n, \beta_k) = P(f_n|S_n)P(q_{nk}|f_n)P(y_n|X_n, q_{nk}, \beta_k). \tag{13}$$

The joint probability of $f_n$ and $y_n$ is then obtained by summing Equation (15) over all classes $K$:

$$\begin{aligned} P(f_n, y_n) &= \sum_{k=1}^{K} P(f_n, y_n, q_{nk}|S_n, X_n, \beta_k) \\ &= \sum_{k=1}^{K} P(f_n|S_n)P(q_{nk}=1|f_n)P(y_n|X_n, q_{nk}=1, \beta_k). \end{aligned} \tag{14}$$

Finally, the joint likelihood function of the GP-LCCM model for a sample of $N$ decision-makers is given by:

$$P(f, y) = \prod_{n=1}^{N} P(f_n, y_n)$$
$$= \prod_{n=1}^{N} \sum_{k=1}^{K} P(f_n|S_n) P(q_{nk} = 1|f_n) \prod_{t=1}^{T_n} \prod_{j=1}^{J} P(y_{njt} = 1|X_{njt}, q_{nk} = 1, \beta_k)^{y_{njt}}. \quad (15)$$

To learn the parameters $\beta_k$ and the hyperparameters of the kernel, the log of the above likelihood should be maximized and evaluated over the unknown parameters. Maximum likelihood estimation techniques are usually used in estimating discrete choice models. However, maximizing the log-likelihood of a discrete choice model with discrete latent variables, such as LCCM, becomes more difficult as the number of classes, and consequently the number of parameters, increases. With larger number of classes and parameters, the calculation of the gradient becomes slower and empirical singularity might arise at some iterations making the inversion of the Hessian matrix numerically challenging (Train, 2008). The Expectation-Maximization (EM) algorithm (Dempster et al., 1977) is a two-stage iterative maximization technique that overcomes the aforementioned complications by repeatedly maximizing an expectation or a lower bound function of the likelihood (Train, 2008). Several studies have applied EM algorithms to discrete choice model applications with mixing distributions or discrete latent variables (Bhat, 1997; El Zarwi, 2017; Sfeir et al., 2020b; Train, 2008, to name a few) and the results showed that such techniques are computationally attractive. The EM algorithm framework consists of two main steps, E (expectation) and M (maximization). In the former (E-step), the expectations of the latent variables conditioned on the current estimates of the unknown parameters and the observed variables are estimated, while in the latter (M-step), the expectation of the log-likelihood is maximized, conditioned on the observed variables and the expectations of the latent variables obtained from the E-step, to update the estimates of the unknown parameters. The algorithm alternates between these two steps until a predefined convergence criterion is satisfied. In this study, an EM-based algorithm is derived and implemented for estimating the proposed GP-LCCM.

*3.4.2 EM Algorithm*

The EM algorithm requires writing the likelihood function (Equation 15) assuming that the class assignments ($q_{nk}$) are no longer latent:

$$P(f, y) = \prod_{n=1}^{N} \prod_{k=1}^{K} [P(f_n|S_n) P(q_{nk} = 1|f_n)]^{q_{nk}}$$
$$\times \prod_{n=1}^{N} \prod_{k=1}^{K} \prod_{t=1}^{T_n} \prod_{j=1}^{J} P(y_{njt} = 1|X_{njt}, q_{nk} = 1, \beta_k)^{y_{njt} q_{nk}}. \quad (16)$$

The logarithm of the above likelihood is then the sum of the two sub-components of the model, the class membership component (first term on the right-hand side below) and the class-specific choice component (second term), as follows:

$$LL = \sum_{n=1}^{N}\sum_{k=1}^{K} q_{nk} log[P(f_n|S_n)P(q_{nk} = 1|f_n)]$$
$$+ \sum_{n=1}^{N}\sum_{k=1}^{K}\sum_{t=1}^{T_n}\sum_{j=1}^{J} y_{njt} q_{nk} log[P(y_{njt} = 1|X_{njt}, q_{nk} = 1, \beta_k)]. \quad (17)$$

Next, the unknown choice model parameters ($\beta_k$) can be estimated by setting the derivatives of the log-likelihood (Equation 17) with respect to the unknown parameters to zero if and only if $q_{nk}$'s are known. Similarly, the hyperparameters of the GP kernel function can be found using the Laplace approximation method if and only if $q_{nk}$'s are known. Therefore, the expectation of $q_{nk}$ (E-step) is calculated using Bayes' theorem as follows:

$$P(q_{nk}|y_n, S_n, f_n, X_n, \beta_k) \propto P(q_{nk}|f_n, S_n) P(y_n|X_n, q_{nk}, \beta_k), \quad (18)$$

$$E[q_{nk}] = \gamma_{q_{nk}} = \frac{P(q_{nk}|f_n, S_n) P(y_n|X_n, q_{nk}, \beta_k)}{\sum_{c=1}^{K} P(q_{nc}|f_n, S_n) P(y_n|X_n, q_{nc}, \beta_c)}. \quad (19)$$

Then, the expected value of the log-likelihood w.r.t $q_{nk}$ is maximized instead of Equation 17, due to the unknown values of the latent variables $q_{nk}$, to find/update the unknown hyper/parameters (M-step). The expected log-likelihood function is given by:

$$E[LL] = \sum_{n=1}^{N}\sum_{k=1}^{K} \gamma_{q_{nk}} log[P(f_n|S_n)P(q_{nk} = 1|f_n)]$$
$$+ \sum_{n=1}^{N}\sum_{k=1}^{K}\sum_{t=1}^{T_n}\sum_{j=1}^{J} y_{njt} \gamma_{q_{nk}} log[P(y_{njt} = 1|X_{njt}, q_{nk} = 1, \beta_k)]. \quad (20)$$

Setting the derivative of the above expected log-likelihood with respect to $\beta_k$ to zero, we can find the updated solution of $\beta_k$ as follows:

$$\beta_k = argmax_{\beta_k} \sum_{n=1}^{N}\sum_{t=1}^{T_n}\sum_{j=1}^{J} y_{njt} \gamma_{q_{nk}} log\left[\frac{e^{X'_{njt}\beta_k}}{\sum_{j'=1}^{J} e^{X'_{nj't}\beta_k}}\right]. \quad (21)$$

Note that closed-form solutions cannot be obtained for Equation 21. Instead, we rely on the gradient-based numerical optimization method BFGS (Nocedal and Wright, 2006) and the constrained version L-BFGS-B (Zhu et al., 1997).

As for estimating the hyperparameters of the Gaussian Process kernel, a Laplace approximation method is applied. However, the target variables (class labels) should be known since Laplace approximation is applied to classification applications. For the sake of simplicity and without loss of generality, we consider the case of two classes ($K = 2$). After calculating the expectations of $q_{nk}$'s in the E-step (which are continuous values between 0 and 1), class labels are generated using hard clustering/assignment as follows: if $E[q_{n0}] > E[q_{n1}]$, then individual $n$ belongs to class 0, otherwise individual $n$ belongs to class 1. Next, a Laplace approximation is applied and the hyperparameters of the kernel are estimated.

The steps of the EM estimation of the proposed GP-LCCM with two classes ($K = 2$) are:

1. Initialize the parameters $\beta_k$ and assign each individual to a class (0 or 1) randomly
2. Select a kernel function and initialize the corresponding hyperparameters
3. **E-step**: Estimate the expectations of $q_{nk}$'s using Equation 19
4. **M-step**:
    i. Re-estimate/update the parameters $\beta_k$ using Equation 21 and the expectations from the E-step
    ii. Assign each individual to one class (0 or 1) using the expectations from the E-step as follows: if $E[q_{n0}] > E[q_{n1}]$, then individual $n$ belongs to class 0, otherwise individual $n$ belongs to class 1
    iii. Re-estimate/update the hyperparameters of the kernel function using the Laplace approximation method and the class labels from the previous step (ii)
5. Evaluate the log-likelihood using the current values of the hyper/parameters and check if the convergence criterion is satisfied. If, not return to step 3
6. Finally, after convergence is reached, estimate the marginal probability of observing a vector of choices $y$, $P(y)$, for all decision-makers $N$ as follows:

$$P(y) = \prod_{n=1}^{N} \sum_{k=1}^{K} P(q_{nk} = 1 | f_n, S_n) P(y_n | X_n, q_{nk} = 1, \beta_k). \tag{22}$$

This marginal probability is calculated in order to compare the GP-LCCM with the traditional LCCM (Equation 7) and assess out-of-sample prediction accuracies.

For multi-class problems ($K = 2$), $K$ binary one-versus-rest classifiers are estimated to classify each class against all the rest using Laplace approximation method. The implementation of the multi-class case is based, within each EM iteration, on Algorithms 3.1, 3.2, and 5.1 from Gaussian Processes for Machine Learning (GPML) by Rasmussen and Williams (2006).

## 4 APPLICATION

We present two different mode choice applications of the proposed GP-LCCM approach. We benchmark the proposed model against the traditional LCCM by using the same specifications for both models. We also compare the results of the first case study to the GBM-LCCM (Sfeir et al., 2020b), which relies on a parametric Gaussian-Bernoulli Mixture model, instead of the non-parametric Gaussian Process, to probabilistically cluster individuals to different classes within a LCCM framework. The GP-LCCM is implemented in Python by using some blocks from: 1) the Gaussian Process Classifier (GPC) of the Scikit-Learn library (Pedregosa et al., 2011), which is based on Laplace approximation by Rasmussen and Williams (2006); 2) and lccm (El Zarwi, 2017a, 2017b), a python package that implements an EM algorithm for estimating traditional latent class choice models. The GBM-LCCM is also implemented in Python using some implementations from the Scikit-learn and lccm packages while the traditional LCCMs are estimated using the lccm package. Convergence of the three different models is supposed to be reached once the change in the log-likelihood function between two successive EM iterations is smaller than $1\times10^{-4}$. All models are estimated five times with different random initialization to assess the stability of the models. All runs are performed on a machine with a core i7 CPU @ 2.40 GHz, 8GB of RAM and a GeForce GT 730M.

## 4.1 Case Study 1: AUB

The dataset used for the first application is from the 2017 Stated Preference (SP) web-based commuting survey of the American University of Beirut (AUB), a private university in Lebanon. The survey was conducted in April of 2017 to investigate the willingness of the AUB population (students, faculty members, and staff) to adopt two new transport modes, shared-taxi and shuttle service, with the aim of mitigating the transportation problems in the surrounding neighborhood that suffers from high level of congestion and parking demand. The proposed shared-taxi is supposed to provide on-demand transport between users' residences and AUB gates (and vice versa). As for the shuttle, it is designed to serve as a non-stop first/last mile service between satellite hubs near the university, where commuters could park their cars, and AUB gates. Information regarding respondents' daily commute to and from AUB, socio-economic/demographic characteristics, and place of residence were collected. Moreover, information about the expected weekly frequencies of using the two proposed modes in addition to their current modes of commute were collected. To do so, four hypothetical scenarios were offered to each respondent where in each scenario the respondent was asked to identify how many weekdays per week she/he is expected to commute using the two new modes in addition to her/his current mode of transport. Figure 2 presents an example of the hypothetical scenarios. For more details regarding the dataset, readers may refer to Sfeir et al. (2020a). For this study, a sub-sample of 650 respondents (corresponding to 2,600 choice observations) of car users who come five days per week to the university is used.

In this application, we model the weekly frequency of commuting by shared-taxi (ST), shuttle (SH), and/or current mode of commute (Car). The latent classes of the three models (LCCM, GBM-LCCM, and GP-LCCM) are characterized by socio-economic variables while the class-specific utility functions of each alternative are characterized by the corresponding travel time, travel cost and constants related to the frequency of using the available modes. Table 1 presents the explanatory variables used in the two components of the three models. Note that the continuous socio-economic variables are standardized (mean = 0 and standard deviation = 1) prior to the estimation of GP-LCCM and GBM-LCCM.

The choice variables are multivariate counts of commuting by three modes during a week with the total count (total number of weekly trips) being fixed to 5 as the number of times an individual commutes to the university is expected to be rather exogenous than endogenous due to institutional constraints on schedule. Multivariate count data with a fixed total count can be modeled by using a full enumeration of all combinations approach (Ben-Akiva and Abou-Zeid, 2013) where the choice set contains all possible combinations of weekly mode frequencies. As such, an alternative is defined as the number of weekly trips an individual would conduct by each of the available modes.

| Shared-Taxi | Shuttle | Your Current Commute to AUB |
|---|---|---|
| **Door-to-door travel time** | **Access travel time** | **Total Travel Time** |
| 33 min | 12 min (by car) | 30 min |
| **Waiting time for late pick-up and/or early drop-off** | **Frequency** | **Mode of Travel** |
| 0 to 5 min | Every 5 min | Car |
| **Number of passengers sharing a ride** | **In-Shuttle travel time** | |
| 4 to 6 (Minivan) | 27 min | |
| **Mobile App/Wi-Fi/Live tracking** | **Wi-Fi/Live tracking** | |
| Available | Not Available | |
| | **One-way shuttle fare including parking cost** | **Parking cost** |
| | 1,000 L.L. | 5,000 L.L. |
| **One-way fare** | **One-way fuel cost (for access by car)** | **One-way fuel cost** |
| 4,000 L.L. | 700 L.L. | 1,800 L.L. |

Based on this scenario, and considering your current pattern to AUB, how many weekdays per week will you use the proposed services? Remember that you indicated earlier that you come on 5 weekdays per week to AUB.

| Shared-Taxi | 0, 1, 2, 3, 4, 5 |
|---|---|
| Shuttle | 0, 1, 2, 3, 4, 5 |
| Your Current Commute to AUB | 0, 1, 2, 3, 4, 5 |

**Figure 2: Example of a hypothetical scenario form the survey (Sfeir et al., 2020a)**

Knowing that three travel modes are available (ST, SH, and Car) and the sample contains people who commute five days per week to AUB, the choice set consists of 21 alternatives. The systematic utility of an individual $n$ choosing a specific sequence of modes $(ST_h, SH_i, Car_j)$ during time period (or scenario) $t$, conditional on her/him belonging to class $k$ can then be specified as follows:

$$\begin{aligned} V_n(ST_h, SH_i, Car_j)_{t|k} &= C_{ST_h,k} + C_{SH_i,k} + C_{Car_j,k} \\ &+ h \times (\beta_{Cost_{ST},k} Cost_{n,ST,t} + \beta_{TT_{ST},k} TT_{n,ST,t}) \\ &+ i \times (\beta_{Cost_{SH},k} Cost_{n,SH,t} + \beta_{TT_{SH},k} TT_{n,SH,t} + \beta_{Head,k} Head_{n,SH,t}) \\ &+ j \times (\beta_{Cost_{Car},k} Cost_{n,Car,t} + \beta_{TT_{Car},k} TT_{n,Car,t}), \end{aligned} \qquad (23)$$

where $h$, $i$, and $j$ are values between 0 and 5 that represent the number of weekday trips by shared-taxi, shuttle and car, respectively. It is assumed that the impact of travel cost, travel time, and headway variables on the utility is proportional to the number of weekly trips by each mode ($h$, $i$, and $j$). Moreover, the travel cost and travel time coefficients are specified as mode-specific. The $C$'s are constants related to the weekly frequency of the three modes and replace the traditional alternative-specific constants (ASCs) that are defined for each alternative (Ben-Akiva and Abou-Zeid, 2013; Sfeir et al., 2020b). Six constants need to be defined for each of the three modes (ST, SH, and Car) since the number of times each mode can be selected per week varies between 0 and 5. Finally, four constants $(C_{ST_0}, C_{SH_0}, C_{Car_0}, C_{Car_5})$ are fixed to zero for identification purposes.

**Table 1: Explanatory variables used in the models**

| Variable | Description | Component |
|---|---|---|
| $Cost_{ST}$ | Cost of a one-way trip by shared-taxi (in 1,000 L.L.)[3] | |
| $Cost_{SH}$ | Cost of a one-way trip by shuttle including parking cost at the satellite parking (in 1,000 L.L.) | |
| $Cost_{Car}$ | Fuel and parking cost of a one-way trip by car (in 1,000 L.L.) | Class-specific choice model |
| $TT_{ST}$ | Travel time of one-way trip by shared taxi (in hours) | |
| $TT_{SH}$ | Travel time of one-way trip by shuttle including access time to the satellite parking (in hours) | |
| $TT_{Car}$ | Travel time of one-way trip by car (in hours) | |
| $Head$ | Shuttle headway (in hours) | |
| $Age$ | Age of the respondent (in years/10) | |
| $Grade$ | A number between 1 and 16 used to specify the job, seniority, and salary of a staff member (Grade/10) | Class membership model |
| $C/D$ | Ratio of number of cars available over number of licensed drivers per household | |
| $Nb$ | Number of people who are usually present in the car during the trip from home to AUB | |

Summary statistics for the LCCM, GBM-LCCM and GP-LCCM are shown in Table 2. LCCM was only able to identify two latent classes. Increasing the number of classes beyond two resulted in class-specific estimates with very high standard errors (identification issues). On the other hand, GBM-LCCM and GP-LCCM were able to identify higher number of classes (up to 5). However, such models with more than three latent classes generated positive travel cost and/or travel time coefficients and as such are excluded from the comparison (Table 2). Knowing that the choice of the kernel function of GPs affects the generalization performance of the model, different kernels and combinations of kernels were tested and models with better generalization performance, better in-sample goodness-of-fit measures (LL, AIC, and BIC), and reasonable parameter estimates signs and magnitudes were selected. For the GP-LCCM with two classes, a Matérn kernel with a smoothness parameter ($v$) of 2.5 was selected while for three latent classes, a combination of a constant and a Matérn kernel with a smoothness parameter of 2.5 iwas used. A constant had to be added since a single Matérn kernel resulted in high standard errors for some class-specific parameter estimates. Similarly, four GMM covariance structures (full, tied, spherical, and diagonal) were tested for the GBM-LCCM and models with better results were selected. A tied covariance structure wherein all latent classes share the same general covariance matrix was used for the case of two classes while a spherical structure wherein each latent class has one single variance was used for the case of three classes. Results show that the GP-LCCM with two classes has better goodness-of-fit measures (Joint LL, LL, AIC, and BIC) and better prediction accuracy in terms of log-likelihood (Pred. LL) than the two other models with two classes. Moreover, the GP-LCCM with three latent classes has better goodness-of-fit and prediction measures than all other models. It is to be noted that, compared to the LCCM with two latent classes, the GP-LCCM with three latent classes improves the in-sample goodness-of-fit (LL) and the out-of-sample prediction accuracy (Pred. LL) by 4.5% and 8.8%, respectively. As for the GBM-LCCM with three latent classes, the improvement is less significant with 0.4% for the LL and 2.6% for the Pred. LL (Table 2). It is believed that the superiority of the GP-LCCM stems from the non-parametric nature of GPs, which allows for more flexibility than the parametric structure of the Gaussian-Bernoulli mixture models and the linear-in-the-parameters utility specification of the class membership model of the traditional LCCM. Note that the 5-fold cross validation technique is used to assess the predictive power of the models. Furthermore, the Akaike Information Criterion (AIC) and the Bayesian Information Criterion (BIC) are derived as follows:

$$AIC = 2M - 2\log LL, \tag{24}$$

$$BIC = M \log D - 2 \log LL, \tag{25}$$

where $M$ is the number of parameters, $LL$ is the estimated marginal choice log-likelihood, and $D$ is the number of data points (observations). For the LCCM and GBM-LCCM, the number of parameters $M$ is equal to the number of unknown parameters that are statistically estimated using the available data. As for the GP-LCCM, the complexity of the model grows with the amount of data used for estimation since Gaussian Processes are non-parametric models. The true number of parameters would be equal to the number of data points (2600 observations for this application) in addition to the number of choice parameters ($\beta_k$) from the class-specific choice models (42 choice parameters for the case of 2 classes). However, such number would be unreasonable for the estimation of AIC and BIC. Instead, it is common to assume that the number of parameters in a GP model is equal to the number of kernel hyperparameters (Lloyd et al., 2014; Richter and Toledano-Ayala, 2015). Therefore, for the sake of comparison with other models, we assume that the number of parameters $M$ for a GP-LCMM is equal to the number of kernel hyperparameters ($\nu$ and $\ell$ in case of a Matérn kernel) in addition to the number of class-specific choice parameters ($\beta_k$).

---
[3] 1 USD = 1,500 Lebanese Lira (L.L.) at the time the survey was conducted.

**Table 2: Summary results of the first application**

| Classes | Model | Number of Hyper/Parameters | Joint LL[a] | LL[b] | AIC | BIC | Average Pred. LL |
|---|---|---|---|---|---|---|---|
| 2 | LCCM | 47 |  | -4,910.92 | 9,915.84 | 10,191.41 | -1,024.93 |
|  | GBM-LCCM | 61 | -8,533.22 | -4,911.08 | 9,944.16 | 10,301.82 | -1,012.62 |
|  | GP-LCCM | 44 | **-4,905.31** | **-4,877.73** | **9,843.46** | **10,101.44** | **-995.76** |
| 3 | GBM-LCCM | 80 | -7,042.21 | -4,893.29 | 9,946.58 | 10,415.64 | -998.41 |
|  | GP-LCCM | 72 | **-4,480.70** | **-4,691.25** | **9,526.50** | **9,948.66** | **-935.23** |

a: joint log-likelihood of the GBM-LCCM and GP-LCCM

b: marginal choice log-likelihood of the LCCM, GBM-LCCM and GP-LCCM

Estimates of the class-specific choice models of the three LCCMs with two latent classes in addition to the corresponding Values of Time (VOTs) are presented in Table 3 (values between parentheses are p-values). Travel time and travel cost parameter estimates have the expected negative sign. In addition, all parameter estimates are similar in magnitude and sign with the ones related to LCCM and GBM-LCCM being almost the same. The second class is characterized by higher VOT for car while individuals belonging to the first class have somewhat close VOTs for car and shuttle, which is also considered a car-based trip since individuals would access the shuttle parking hubs by car. Table 4 shows the estimates of the class-specific choice models and VOTs of GBM-LCCM and GP-LCCM with three latent classes. All travel time and travel cost coefficients have the same order of magnitude and expected negative sign. As for the constants related to the travel mode frequencies ($C's$), some differences can be marked especially for the third class. Moreover, according to GBM-LCCM, individuals belonging to the third class seem to be insensitive towards travel cost of car (p-value = 0.11) and travel time of shuttle (p-value = 0.26) while the same coefficients are highly significant according to GP-LCCM (p-values = 0). This coefficients' insignificance in the third class of GBM-LCCM results in very high and low VOTs of car and shuttle, respectively. However, compared to previous studies (Al-Ayyash et al., 2016; Sfeir et al., 2020a), GP-LCCM generates more consistent estimates of Values of Time than the GBM-LCCM. This previous discussion shows that the behavioral and economic interpretability of the class-specific choice models were not jeopardized by the introduction of Gaussian Processes to the LCCM framework. Furthermore, GP-LCCM is capable of improving the prediction accuracy, capturing more complex heterogeneity than LCCM since a higher number of classes is identified, and generating more reliable VOT estimates than GBM-LCCM.

Results of the class membership model of the LCCM and GBM-LCCM with two latent classes are presented in Table 5. According to both the LCCM parameter estimates and the GBM-LCCM mean estimates, members of the first class are more likely to be older people and staff with high grades who belong to households with high car ownership and have tendency to share rides to AUB. As for the GP-LCCM, the introduction of the non-parametric Gaussian Processes makes the model less transparent at the class membership level. However, the latent classes can still be interpreted although in a different manner than in traditional DCMs and parametric models where interpretability is based on the parameter estimates. Recently, interpretability of machine learning models has become a fundamental area of research and many studies have shown that different techniques can be used for model interpretation (Doshi-velez and Kim, 2017; Ribeiro et al., 2016a, 2016b; Wang et al., 2020b, to name a few). One approach to interpret "black box" machine learning models is using model-agnostic techniques that infer explanations from the estimated/trained model by treating it as a black box (Ribeiro et al., 2016b). In this application, we rely on the Local Interpretable Model-agnostic Explanations (LIME) technique to interpret the class membership component of the GP-LCCM. LIME (Ribeiro et al., 2016a) learns an interpretable model on top of the original machine learning model with the aim of interpreting individual predictions. First, LIME generates a new dataset by shuffling the original observations (the socio-economic variables used for clustering). Second, LIME weights the new observations by their closeness to the original dataset. Finally, LIME fits an interpretable model (e.g. linear regression) by using the new shuffled-weighted observations and their associated predictions (class labels) from the original model (GP-LCCM). Figure 3 explains the class predictions of the GP-LCCM with two latent classes. Several individual predictions have been investigated. However, for the sake of brevity, only three observations are presented. The bar charts in Figure 3 portray the importance of each variable, with the value next to each bar being the corresponding weight, while the colors (blue for class 1 and orange for class 2) specify which class the variables contribute to. In the three bar charts of Figure 3, LIME assigns positive and blue weights to Age and Grade while the weights of C/D and Nb are close to zero. This implies that the first class (blue color) is characterized by higher age and grade values which is in line with the corresponding positive parameters from the LCCM and GBM-LCCM (Table 5). Moreover, the order of magnitude of the weights, which represent the importance of each variable, is similar to the order of parameters' magnitude from the LCCM and GBM-LCCM. Both models from Table 5 have high parameter/mean estimates for Age and Grade while the ones related to C/D and Nb are lower and insignificant according to the LCCM (high p-values). Finally, the first two individuals (Figures 3.a and 3.b) are more likely to belong to the second class (class probabilities higher than 0.5) while the third individual (Figure 3.c) is more likely to belong to the first class (class probability = 0.75). This shows that although Gaussian Processes make the class components less transparent, local interpretability can still be achieved by relying on model-agnostic techniques. It is to be noted that local interpretability is sometimes much more relevant for model explainability than abstract global interpretation techniques (Montavon et al., 2018), especially given that the former targets individual explanations that can offer better in-depth realization of features contribution/importance in smaller groups of individuals (Kopitar et al., 2019).

As for the computational times, LCCM and GBM-LCCM take on average less than a minute to converge while the GP-LCCMs with two and three classes take around 5 and 35 minutes, respectively. This difference in runtime is expected due to the non-parametric nature of GPs whose inference requires inverting the covariance matrix with a cost of scale $O(N^3)$. However, it is to be noted that all runs were performed on a machine with a single core, meaning the implementation was not optimized to take full advantage of modern computational hardware (e.g. GPU) which could make computational overheads less relevant.

Table 3: Class-Specific Choice Models and VOT (K = 2)

| Variable | LCCM | | GBM-LCCM | | GP-LCCM | |
|---|---|---|---|---|---|---|
| | Class 1 | Class 2 | Class 1 | Class 2 | Class 1 | Class 2 |
| | | | Class-specific choice model | | | |
| $C_{Car1}$ | -2.56 (0.00) | 0.372 (0.00) | -2.50 (0.00) | 0.361 (0.00) | -2.69 (0.00) | 0.351 (0.00) |
| $C_{Car2}$ | -2.06 (0.00) | 0.298 (0.01) | -2.04 (0.00) | 0.290 (0.01) | -1.92 (0.00) | 0.282 (0.01) |
| $C_{Car3}$ | -2.36 (0.00) | 0.516 (0.00) | -2.39 (0.00) | 0.508 (0.00) | -2.64 (0.00) | 0.483 (0.00) |
| $C_{Car4}$ | -3.09 (0.00) | -0.422 (0.01) | -3.08 (0.00) | -0.430 (0.00) | -2.77 (0.00) | -0.449 (0.00) |
| $C_{ST1}$ | -1.60 (0.03) | -0.464 (0.00) | -1.62 (0.04) | -0.465 (0.00) | -1.96 (0.00) | -0.475 (0.00) |
| $C_{ST2}$ | -2.10 (0.00) | -0.172 (0.24) | -2.09 (0.00) | -0.174 (0.24) | -1.83 (0.00) | -0.211 (0.15) |
| $C_{ST3}$ | -1.05 (0.03) | -0.108 (0.61) | -1.08 (0.03) | -0.108 (0.61) | -1.46 (0.02) | -0.0979 (0.63) |
| $C_{ST4}$ | -3.08 (0.00) | -0.347 (0.25) | -3.15 (0.00) | -0.347 (0.25) | -2.24 (0.03) | -0.434 (0.15) |
| $C_{ST5}$ | -0.158 (0.53) | -0.217 (0.53) | -0.159 (0.53) | -0.209 (0.55) | -0.0556 (0.83) | -0.169 (0.62) |
| $C_{SH1}$ | -2.26 (0.00) | -0.280 (0.02) | -2.30 (0.00) | -0.286 (0.02) | -2.83 (0.00) | -0.326 (0.01) |
| $C_{SH2}$ | -2.99 (0.00) | 0.413 (0.00) | -3.03 (0.00) | 0.403 (0.00) | -2.86 (0.00) | 0.350 (0.01) |
| $C_{SH3}$ | -2.26 (0.00) | 0.678 (0.00) | -2.29 (0.00) | 0.661 (0.00) | -2.47 (0.00) | 0.595 (0.00) |
| $C_{SH4}$ | -3.93 (0.00) | 0.373 (0.09) | -4.02 (0.00) | 0.354 (0.11) | -3.95 (0.00) | 0.282 (0.19) |
| $C_{SH5}$ | -1.52 (0.00) | 0.378 (0.16) | -1.52 (0.00) | 0.379 (0.15) | -1.54 (0.00) | 0.367 (0.15) |
| $\beta_{Cost\_Car}$ | -0.0446 (0.00) | -0.0456 (0.00) | -0.0442 (0.00) | -0.0462 (0.00) | -0.0425 (0.00) | -0.630 (0.00) |
| $\beta_{Cost\_ST}$ | -0.101 (0.00) | -0.109 (0.00) | -0.101 (0.00) | -0.110 (0.00) | -0.105 (0.00) | -0.0474 (0.00) |
| $\beta_{Cost\_SH}$ | -0.0400 (0.00) | -0.0998 (0.00) | -0.0401 (0.00) | -0.0993 (0.00) | -0.0399 (0.00) | -0.629 (0.00) |
| $\beta_{Time\_Car}$ | -0.409 (0.00) | -0.658 (0.00) | -0.409 (0.00) | -0.653 (0.00) | -0.420 (0.00) | -0.108 (0.00) |
| $\beta_{Time\_ST}$ | -0.372 (0.00) | -0.646 (0.00) | -0.372 (0.00) | -0.641 (0.00) | -0.380 (0.00) | -0.375 (0.00) |
| $\beta_{Time\_SH}$ | -0.252 (0.00) | -0.387 (0.00) | -0.252 (0.00) | -0.384 (0.00) | -0.255 (0.00) | -0.0971 (0.00) |
| $\beta_{Head}$ | -0.0423 (0.65) | -0.565 (0.00) | -0.0442 (0.64) | -0.561 (0.00) | -0.0556 (0.56) | -0.562 (0.00) |
| | | | $VOT$ ($/hr$) | | | |
| $Car$ | 6.11 | 9.61 | 6.16 | 9.42 | 6.59 | 8.87 |
| $ST$ | 2.44 | 3.96 | 2.45 | 3.90 | 2.42 | 3.87 |
| $SH$ | 4.20 | 2.59 | 4.19 | 2.58 | 4.26 | 2.58 |

Table 4: Class-Specific Choice Models and VOT (K = 3)

| Variable | GBM-LCCM | | | GP-LCCM | | |
|---|---|---|---|---|---|---|
| | Class 1 | Class 2 | Class 3 | Class 1 | Class 2 | Class 3 |
| | Class-specific choice model | | | | | |
| $C_{Car1}$ | -2.24 (0.00) | 0.444 (0.00) | -0.102 (0.78) | -3.54 (0.29) | 0.746 (0.33) | 0.397 (0.00) |
| $C_{Car2}$ | -1.82 (0.00) | 0.290 (0.02) | 0.327 (0.25) | -2.83 (0.57) | 1.56 (0.01) | -0.0479 (0.77) |
| $C_{Car3}$ | -2.11 (0.00) | 0.677 (0.00) | -0.0412 (0.88) | -2.52 (0.62) | 1.90 (0.00) | 0.140 (0.48) |
| $C_{Car4}$ | -2.90 (0.00) | -0.224 (0.18) | -1.26 (0.00) | -5.20 (0.20) | 0.347 (0.16) | -1.34 (0.00) |
| $C_{ST1}$ | -1.61 (0.01) | -0.331 (0.01) | -1.11 (0.00) | -1.33 (0.73) | -1.45 (0.00) | 0.285 (0.07) |
| $C_{ST2}$ | -2.31 (0.00) | -0.00930 (0.95) | -0.907 (0.03) | -2.85 (0.57) | -1.90 (0.00) | 0.915 (0.00) |
| $C_{ST3}$ | -1.07 (0.02) | 0.0262 (0.91) | -0.938 (0.09) | -1.40 (0.78) | -1.41 (0.03) | 1.09 (0.00) |
| $C_{ST4}$ | -3.23 (0.00) | -0.073 (0.82) | -1.78 (0.08) | -1.71 (0.61) | -1.06 (0.23) | 0.863 (0.05) |
| $C_{ST5}$ | -0.181 (0.47) | -0.284 (0.48) | -0.244 (0.76) | -0.141 (0.60) | -1.20 (0.06) | 2.69 (0.00) |
| $C_{SH1}$ | -2.19 (0.00) | -0.240 (0.07) | -0.557 (0.08) | -3.59 (0.267) | -0.838 (0.00) | 0.987 (0.00) |
| $C_{SH2}$ | -3.17 (0.00) | 0.556 (0.00) | -0.535 (0.11) | -3.48 (0.49) | -0.812 (0.08) | 2.08 (0.00) |
| $C_{SH3}$ | -2.12 (0.00) | 0.862 (0.00) | -0.959 (0.03) | -1.84 (0.71) | 0.285 (0.68) | 2.18 (0.00) |
| $C_{SH4}$ | -4.38 (0.00) | 0.627 (0.01) | -1.63 (0.00) | -4.36 (0.27) | -0.231 (0.80) | 2.45 (0.00) |
| $C_{SH5}$ | -1.45 (0.00) | 0.601 (0.04) | -1.15 (0.05) | -1.64 (0.00) | 0.555 (0.35) | 3.17 (0.00) |
| $\beta_{Cost\_Car}$ | -0.0451 (0.00) | -0.0707 (0.00) | -0.0172 (0.11) | -0.0451 (0.00) | -0.0524 (0.00) | -0.0645 (0.00) |
| $\beta_{Cost\_ST}$ | -0.0991 (0.00) | -0.107 (0.00) | -0.123 (0.00) | -0.102 (0.00) | -0.101 (0.00) | -0.132 (0.00) |
| $\beta_{Cost\_SH}$ | -0.0421 (0.00) | -0.0832 (0.00) | -0.120 (0.00) | -0.0411 (0.00) | -0.150 (0.00) | -0.118 (0.00) |
| $\beta_{Time\_Car}$ | -0.410 (0.00) | -0.717 (0.00) | -0.614 (0.00) | -0.404 (0.00) | -0.680 (0.00) | -0.466 (0.00) |
| $\beta_{Time\_ST}$ | -0.384 (0.00) | -0.777 (0.00) | -0.349 (0.01) | -0.360 (0.00) | -0.856 (0.00) | -0.619 (0.00) |
| $\beta_{Time\_SH}$ | -0.259 (0.00) | -0.519 (0.00) | -0.107 (0.26) | -0.226 (0.00) | -0.752 (0.00) | -0.319 (0.00) |
| $\beta_{Head}$ | -0.0114 (0.91) | -0.757 (0.00) | -0.380 (0.06) | -0.0551 (0.59) | -0.783 (0.00) | -0.381 (0.00) |
| | $VOT$ ($/hr$) | | | | | |
| $Car$ | 6.07 | 6.76 | 23.81 | 5.96 | 8.65 | 4.82 |
| $ST$ | 2.58 | 4.85 | 1.89 | 2.34 | 5.64 | 3.13 |
| $SH$ | 4.10 | 4.16 | 0.60 | 3.67 | 3.35 | 1.80 |

Table 5: Class membership estimates of LCCM and GBM-LCCM (K = 2)

| | LCCM | | | GBM-LCCM | |
|---|---|---|---|---|---|
| | Class 1 | Class 2 | | Class 1 | Class 2 |
| $ASC$ | -2.271 (0.00) | - | $\pi$ | 0.575 | 0.425 |
| $\beta_{Age}$ | 0.587 (0.00) | - | $\mu_{Age}$ | 0.303 | -0.409 |
| $\beta_{Grade}$ | 0.569 (0.00) | - | $\mu_{Grade}$ | 0.225 | -0.303 |
| $\beta_{C/D}$ | 0.267 (0.37) | - | $\mu_{C/D}$ | 0.0459 | -0.062 |
| $\beta_{Nb}$ | 0.0850 (0.26) | - | $\mu_{Nb}$ | 0.0513 | -0.0693 |

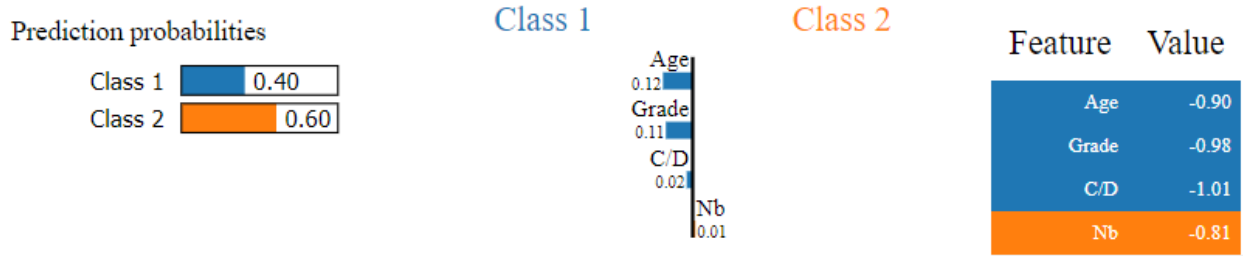
a) Individual 1: Age = 24, Grade = 0, C/D = 0.5, Nb = 0

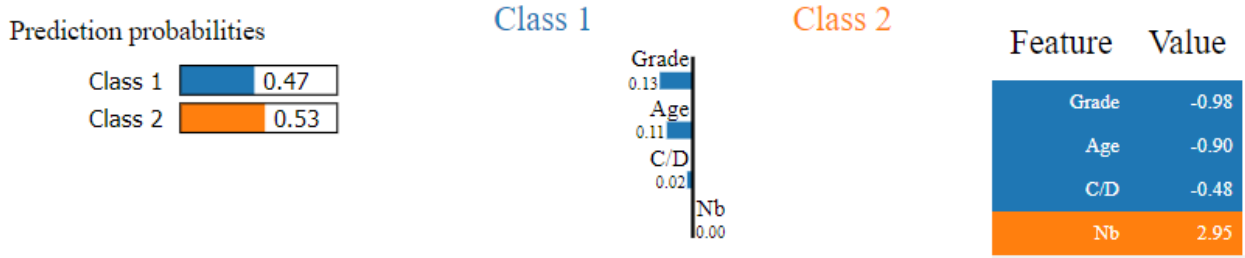
b) Individual 2: Age = 24, Grade = 0, C/D = 0.67, Nb = 5

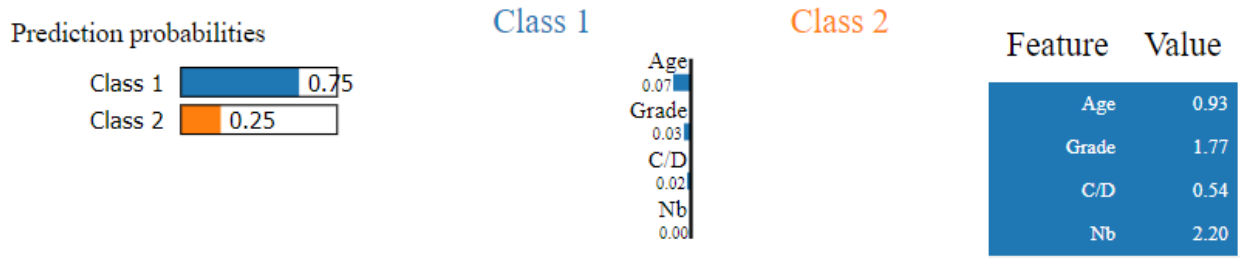
c) Individual 3: Age = 45, Grade = 16, C/D = 1, Nb = 4

**Figure 3: Explaining three individual class predictions of the GP-LCCM with two classes (K = 2) using LIME**

## 4.2 Case Study 2: Swissmetro

The second application is based on the famous Swissmetro dataset which consists of SP survey data collected in Switzerland during March of 1998 to assess the potential demand for the formerly proposed Swissmetro, a maglev underground transport system (Bierlaire et al., 2001). Each respondent was offered 9 hypothetical scenarios with the following alternatives: Train, Swissmetro (SM) and Car (only for car owners). Each alternative was described by its corresponding attributes such as travel time and travel cost/fare, etc. Socio-economic/demographic information was also collected. For more information, readers may refer to (Bierlaire, 2018; Bierlaire et al., 2001). The original dataset contains 10,728 observations corresponding to 1,192 respondents (751 car users and 441 rail-based travelers). However, observations with missing age, unknown choices and "other" trip purposes are removed. As a result, the used sample consists of 10,692 observations corresponding to 1,188 respondents. The sample is randomly divided into 80% (950 respondents and 8,550 observations) for training/estimation and 20% (238 respondents and 2,142 observations) for testing/prediction.

In this application, we only compare GP-LCCM to LCCM. GBM-LCCM is not considered since no continuous variables are used for clustering. The latent classes of the two models are characterized by the categorical variables AGE, MALE, INCOME, FIRST, LUGGAGE and PURPOSE as shown in Table 6. As for the class-specific choice models, the utilities of the three alternatives are specified using generic travel time and travel cost coefficients in addition to alternative-specific constants for the Train and Car alternatives. We make use of the L-BFGS-B optimizer (Zhu et al., 1997) to constrain the signs of the travel time and travel cost parameters since both models generated counter-intuitive positive signs for travel cost and/or travel time coefficients when using the unbounded optimizer BFGS (Nocedal and Wright, 2006). The number of latent classes is varied from 2 to 10 and the models are estimated 5 times with different random initializations to assess the stability of the models. All LCCMs (K = 2 to 5) take less than a minute to converge, while the computational time of GP-LCCMs varies between 5 and 55 minutes.

Table 7 is dedicated to the summary statistics of LCCMs and as such shows the Log-Likelihood (LL), Akaike Information Criterion (AIC), Bayesian Information Criterion (BIC) and log-likelihood of the test sample (Pred. LL). Increasing the number of latent classes beyond 5 for LCCM resulted in some travel time and travel cost parameters with a zero value while other parameter estimates from both sub-components had very high standard errors. It is clear that 5 is the optimal number of classes for LCCM since the corresponding model has the lowest LL, AIC, BIC and Pred. LL. Table 8 presents the same measures shown in Table 7 in addition to the joint LL of the GP-LCCMs. Estimating models with 8 or more classes generated zero values for some of the constrained parameters (travel time and/or travel cost). Similarly to the previous case study, a manual search was conducted to find the optimal kernel function or combination of kernels. Consequently, a Matérn kernel with a smoothness parameter ($\nu$) of 1.5 was used for all GP-LCCMs. Results show that the 7-class model has the lowest joint LL, LL, AIC, BIC and Pred. LL. Compared to the GP-LCCMs with the same number of classes, LCCMs have better LL (for $K = 4$ and 5) and better prediction accuracy (for $K = 2, 3, 4$ and 5). However, the optimal GP-LCCM with 7 classes outperforms the optimal LCCM with 5 classes over all statistical measures. Results show that the proposed GP-LCCM has the ability to improve the representation of unobserved heterogeneity by identifying a higher number of latent classes, thus improving the model fit and generalization performance. Compared to the best LCCM (K=5), the best GP-LCCM (K=7) improves the in-sample goodness-of-fit (LL) and the prediction accuracy (Pred. LL) by 1% and 2%, respectively. Finally, the 7-class GP-LCCM generates VOTs between 0.01 and 8.96 CHF/min. In a previous study by Han (2019), a nonlinear LCCM with 6 classes generated VOTs between 0.03 and 6.86 CHF/min while Bierlaire et al. (2001) showed, using different MNL and Nested Logit specifications, that the VOT is around 1.2 CHF/min. Given that a relatively high number of classes (seven) is estimated, it is expected to have few classes with high or low Values of Time due to the insignificance of some parameter estimates of travel time and/or travel cost.

**Table 6: Variables used to define the latent classes**

| Variable | Description | Levels |
|---|---|---|
| AGE | The age class of respondents | Age ≤ 24*; 24 < Age ≤ 39; 39 < Age ≤ 54; 54 < Age ≤ 65; Age > 65 |
| MALE | The respondent's gender | 1: Male; 0: Female |
| INCOME | The respondent's income per thousand CHF per year | INCOME < 50*; 50 ≤ INCOME ≤ 100; INCOME > 100; M_INCOME: unknown income |
| FIRST | First class traveler | 0: no; 1: yes |
| LUGGAGE | Number of luggage the respondent carries during a trip | 0: none; 1: one piece; 2: more than one piece* |
| PURPOSE | Purpose of the trip | 1: Commuter; 2: Shopping; 3: Business; 4: Leisure* |

*: level kept as a base

**Table 7: LCCM**

| K | Nb of Parameters | LL | AIC | BIC | Pred. LL |
|---|---|---|---|---|---|
| 2 | 23 | -5,930.76 | 11,907.52 | 12,069.75 | -1,490.62 |
| 3 | 42 | -5,202.71 | 10,489.41 | 10,785.67 | -1,329.94 |
| 4 | 61 | -4,870.51 | 9,863.02 | 10,293.29 | -1,245.18 |
| 5 | 80 | -4,687.99 | 9,535.99 | 10,100.28 | -1,233.69 |

**Table 8: GP-LCCM**

| K | Nb of Parameters | Joint LL[a] | LL[b] | AIC | BIC | Pred. LL |
|---|---|---|---|---|---|---|
| 2 | 10 | -5,930.02 | -5,916.43 | 11,852.86 | 11,923.39 | -1,493.52 |
| 3 | 18 | -4,879.78 | -5,176.06 | 10,388.11 | 10,515.08 | -1,354.44 |
| 4 | 24 | -4,260.21 | -4,878.84 | 9,805.68 | 9,974.97 | -1,263.21 |
| 5 | 30 | -3,872.87 | -4,825.55 | 9,711.11 | 9,922.72 | -1,256.62 |
| 6 | 36 | -3,564.44 | -4,742.13 | 9,556.26 | 9,810.19 | -1,237.09 |
| **7** | **42** | **-3,346.24** | **-4,649.20** | **9,382.39** | **9,678.65** | **-1,213.44** |

a: joint log-likelihood of the GP-LCCM

b: marginal log-likelihood of the GP-LCCM

# 5 CONCLUSION

This paper integrates non-parametric probabilistic machine learning into discrete choice models to improve representation of unobserved heterogeneity. Specifically, we contribute to the literature by presenting a new choice model that combines Gaussian Processes and random utility models within a LCCM framework. The resulting model is a Gaussian Process – Latent Class Choice Model (GP-LCCM) that, similarly to traditional LCCMs, consists of two components, a class membership model and a class-specific choice model. The former is constructed as a Gaussian Process to model unobserved heterogeneity as a discrete construct (latent classes) while the latter estimates the corresponding choice probabilities using logit models. Moreover, an EM algorithm is derived and implemented to estimate/infer the parameters of the choice models and the hyperparameters of the GP kernel function. The iterative nature of the EM algorithm enables the use of the Laplace approximation method to infer the GP posterior for clustering purposes. The model is tested on two different mode choice applications and benchmarked against LCCM and Gaussian Bernoulli Mixture – Latent Class Choice Models (GBM-LCCMs). The findings indicate that the GP-LCCM allows for a higher degree of flexibility by estimating more latent classes than the benchmark models. Moreover, it is capable of improving the in-sample goodness-of-fit measures and the out-of-sample predictive power. This is due to the fact that GPs rely on a non-parametric structure that lessens the restrictive parametric assumptions of the GBM-LCCM and allows more flexibility than the linear specifications of the class membership utilities of traditional LCCMs. Results also showed that the use of Gaussian Processes did not jeopardize the economic and behavioral interpretation of the class-specific choice models. In fact, marginal effects and economic indicators such as VOTs can be easily derived from the model.

Three limitations can be identified. First, the interpretation of the latent classes becomes less transparent. However, latent classes can still be interpreted locally by means of model-agnostic techniques such as LIME. Second, the use of GPs places additional burden on the modeler to select an appropriate kernel function or a combination of kernel functions. Future work could explore ways to automate this task by automatically searching for the kernel structure that would maximize the marginal choice log-likelihood of the overall model. Third, the non-parametric nature of GPs make the estimation process computationally expensive, especially for large datasets. Such limitation could be overcome by using Sparse Gaussian Processes that significantly reduce the time complexity (Titsias, 2009). Finally, although two different mode choice applications have been considered in this study, the proposed model should be applied to different type of datasets (e.g. stated preference and revealed preference) to examine whether the same findings could be reached or not.

## Acknowledgments

This research did not receive any specific grant from funding agencies in the public, commercial, or not-for-profit sectors. However, the first author received funding for his research from the Maroun Semaan Faculty of Engineering and Architecture at the American University of Beirut (AUB).